\documentclass[12pt]{article}
\usepackage{epsfig}

\parskip=\medskipamount 
plus.3em minus.5em \abovedisplayshortskip=.5em plus.2em minus.4em
\renewcommand{\thesection}{\arabic{section}}

\catcode`@=11

\@addtoreset{equation}{section} \@addtoreset{equation}{subsection}
\def\theequation{\ifnum\value{section}=0 \arabic{equation}\ignorespaces
\else \ifnum\value{section}=-1 A.\arabic{equation}\ignorespaces
\else \ifnum\value{subsection}=0
\thesection.\arabic{equation}\ignorespaces \else
\thesection.\arabic{subsection}.\arabic{equation}\ignorespaces
                             \fi
                        \fi
                   \fi}

{\catcode`\'=\active \def'{{}^\bgroup\prim@s}}

\catcode`@=12

\newcommand{\bq}{\begin{equation}}
\newcommand{\be}{\begin{equation}}
\newcommand{\fq}{\end{equation}}
\newcommand{\ee}{\end{equation}}
\newcommand{\bqr}{\begin{eqnarray}}
\newcommand{\beqs}{\begin{eqnarray}}
\newcommand{\fqr}{\end{eqnarray}}
\newcommand{\eeqs}{\end{eqnarray}}

\newcommand{\rf}[1]{(\ref{#1})}

\def\bop#1{\setbox0=\hbox{$#1M$}\mkern1.5mu
    \vbox{\hrule height0pt depth.04\ht0
    \hbox{\vrule width.04\ht0 height.9\ht0 \kern.9\ht0
    \vrule width.04\ht0}\hrule height.04\ht0}\mkern1.5mu}
\def\Box{{\mathpalette\bop{}}}                        

\begin{document}
\thispagestyle{empty}

\begin{flushright}
\begin{tabular}{l}
hep-th/0506013 \\
\end{tabular}
\end{flushright}

\vskip .6in
\begin{center}

{\bf  A Map from Scalar Field Theory to Integer Polynomial Solutions}

\vskip .6in

{\bf Gordon Chalmers}
\\[5mm]

{e-mail: gordon@quartz.shango.com}

\vskip .5in minus .2in

{\bf Abstract}

\end{center}

The terms in the quantum scattering in scalar field theory models is 
parameterized by the invariants $\prod s_{ij}^{n_{ij}}$.  The $s_{ij}$ 
are kinematic two-particle invariants, and the $n_{ij}$ are integers.  
The coefficients of these terms are computed via integrating all Feynman 
diagrams, or within the derivative expansion by solving the iteration 
equations.  The latter has been provided recently; the functions 
which are prefactors of the individual terms $\prod s_{ij}^{n_{ij}}$ 
can be interpreted as terms in the expansions of L-series, which may 
be specified by collections of their zeroes.  Once finding the 
appropriate elliptic curve coefficients, these quantum field solutions 
provide an algorithm to determining all of the mod p zeros to the 
algebraic curves.  The latter is presumably determined by 'experimental' 
computer modeling or by the appropriate determination of the quantum 
prefactors.  

\vfill\pagebreak

The quantum solution to scalar field models is determined by the initial 
conditions of the bare Lagrangian.  Given the initial conditions, there 
is an algorithm for determining the quantum scattering.  These initial 
conditions are examined, together with solving all of the integrals, 
for $\phi^3$ theory in \cite{ChalmersOne}.   The solutions to the terms 
in the effective action, or scattering, is developed in prior works.

The bare Lagrangian is contained in the expansion, 

\bqr 
{\cal L} = {1\over 2} \phi\Box\phi + {1\over 2}m^2 \phi^2 + {1\over 3!} 
 \lambda_3 \phi^3  
\label{phi3expansion} 
\fqr 
with the higher order terms 

\bqr 
{\cal L}_{\rm bare} = {1\over 4!} \phi^4 + \sum {\cal O}_{i} \ . 
\label{higherorder} 
\fqr 
The operators ${\cal O}_{i}$ may contain terms such as $\Box \phi^2 
\Box \phi^2$ and all combinations of derivatives with the fields $\phi$.  
The initial theory in \rf{phi3expansion} and \rf{higherorder}.  

The solution to these scalar models requires the summation of terms, 
i.e. rainbow graphs, illustrated in \cite{ChalmersOne}.  These graphs 
were integrated in \cite{ChalmersOne}, and the integration leads to 
a simplification of the arbitrary loop graphs in the conventional 
Feynman diagrams.  The summation appears complicated due to the 
combinatorics of the tree diagrams obtained from the theory in 
\rf{phi3expansion} and \rf{higherorder}; however, there are simplifications 
that allow the summation to be performed \cite{ChalmersTwo} in the 
arbitrary bare Lagrangian described.  

The scattering at $n$-point in these {\it scalar} models is described 
by functions in $k$-space, 

\bqr 
A_{s_{ij}} = f_{s_{ij},n_{ij}} \prod s_{ij}^{n_{ij}} \ , 
\fqr 
with the invariants $s_{ij}=(k_i+k_j)$ and with the $n_{ij}$ integers 
labeling the expansion.  The set of all combinations spans the complete 
scattering at an arbitrary $n$-point (with $n$ labeling the $k_i$ in 
the product).  The functions $f$ are functions of the couplings $\lambda_3$, 
$\lambda_4$, as well as the coefficients of the higher derivative terms in 
the bare theory as spanned by the operators ${\cal O}_i$.   

The determination of the functions $f_{s_{ij},n_{ij}}$ are given in 
\cite{ChalmersOne}.  It is remarkable that a closed form solution can 
almost be given, in \cite{ChalmersOne}, for an arbitrary bare Lagrangian.  
(As mentioned, the summations in \cite{ChalmersOne} can also be obtained, 
most likely, in \cite{ChalmersTwo}.)  

The determination of the functions $f_{s_{ij},n_{ij}}$ could require a 
different determination, for example requiring the period matrix as in 
the case of the $N=2$ superpotential (or prepotential).  However, in 
the scalar field models there are an infinite number of functions spanned 
by the non-two derivative terms.   

These functions are a power series in the coupling $\lambda_3$, and in 
the other couplings.  If all of the couplings are chosen to be proportional 
to $\lambda_3$, with factors of the mass $m$ (or in a conventional cutoff 
setting involving $\Lambda$), then these functions are arbitrary.  This 
would involve power series possibly multiplying the bare terms.   However, 
the theory involving only the $\lambda_3$ term is sufficient for this 
analysis.   

The series of functions $f_{s_{ij},n_{ij}}$ span a set of determined 
functions, all of which in perturbation theory are power series in the 
coupling $\lambda$ (i.e. in $\phi^3$, $\phi^4$, or with the improvements).  

These functions can be turned into a collection of zeros.  For example, 
these functions, as 

\bqr 
f_i = \sum a_{i,j} \lambda^j 
\fqr 
can be converted into the zero set, 

\bqr 
f_i = \prod (\tilde\lambda_{i,a}+\lambda_a)  
\label{zeroset} 
\fqr 
with the factors $\tilde\lambda_{i,a}$ complex variables.  The $i$ 
labels the various terms in the quantum scattering at $n$-point.  These 
functions can be interpreted as the zero set to an associated L-series 

\bqr 
L(C,\lambda) = \prod (1-a_p p^{-2\lambda} + p^{1-2\lambda})^{-1} \ . 
\label{Lseries} 
\fqr 
These L-series have an arbitary set of zeros on the complex plane, 
given the parameters $a_p=N_p+p$ which might be arbitrary; however, the 
curve and its solutions 

\bqr 
y^2=x^3+a x + b ~{\rm mod~ p}  
\fqr  
with $p$ prime generate these solutions; $N_p$ counts the solutions 
at a given prime $p$.  The numbers $a$ and $b$ which 
determine these solutions could be large or small.  These L-series 
have an expansion, in $\lambda$ which is a power series; they also 
are described by the zero set of the function as in \rf{zeroset} 
(the simplest example is the Riemann zeta function).   

The interpretation of the quantum scattering, as described in the scalar 
field theory \cite{ChalmersOne} and in \cite{ChalmersTwo}, in terms of the 
elliptic function is valuable for a number of reasons.  Primarily, once 
the curves' numbers $a_i$ and $b_i$ are determined, then all of the 
solutions' countings $a_p$ in \rf{Lseries} can be found by an inverse 
transform; the solution to the $a_i$ and $b_i$ curve parameters depends 
on the $n$-point number and the set of integers $n_{ij}$ (defining $\prod 
s_{ij}^{n_{ij}}$).  Given the parameters $a_i$ and $b_i$, this solution can 
be obtained, of which there no known recipe is known.   

The terms in the effective action has been explained for an arbitrary 
tree solution in \cite{ChalmersOne} and \cite{ChalmersTwo}.  Apart from 
a solution based on the integrated terms as generated in a general 
setting \cite{ChalmersOne}, the symmetries of these tree diagrams is 
important to find the parameters $a_i$ and $b_i$ for the general 
term $\prod s_{ij}^{n_{ij}}$.  

The parameters $a_i$ and $b_i$ require a different determination of 
the functions $f_i$ from the explicit integral solution based on the 
derivative expansion.   Once these terms are described, for a potential 
arbitrary set of conditions in the bare theory, which would generate 
an arbitrary curve (with $a_i$ and $b_i$), then any elliptic curve 
could be modeled.  This elliptic curve modeling would generate all 
of the solution numbers $a_p$ in \rf{Lseries}.  

This work has bearing on the base interpetration of the elliptic 
modeling in \cite{ChalmersThree}, the gauge and gravity quantum solution 
(with \cite{ChalmersFour}), and the compact expressions for gauge 
amplitudes in \cite{ChalmersFive}.

\vfill\break


\begin{thebibliography}{99}

\bibitem{ChalmersOne}
G. Chalmers, {\it Quantum Solutions to Scalar Field Theory Models}, 
physics/050518.

\bibitem{ChalmersTwo} 
G. Chalmers, {\it in progress}.  

\bibitem{ChalmersThree} 
G. Chalmers, {\it Integer and Rational Solutions to Polynomial Equations}, 
physics/0503200.  

\bibitem{ChalmersFour} 
G. Chalmers, {\it Quantum Gauge Theory Amplitude Solutions}, 
physics/0505077.

\bibitem{ChalmersFive} 
G. Chalmers, {\it Very Compact Expressions for Amplitudes}, physics/0502058. 

\end{thebibliography}
\end{document}